\documentclass[12pt,a4paper]{article}
\usepackage{tikz}
\usepackage{authblk}
\usepackage{amsmath,amssymb}
\usepackage{graphicx}
\usepackage{amsfonts}
\usepackage{epsfig}
\usepackage{times,pstricks,pst-node}
\usepackage{pst-all} 
\usepackage{tikz}
\usepackage{pgflibrarysnakes}
\usetikzlibrary[snakes,arrows,automata,backgrounds,calendar,chains,matrix,mindmap,patterns,petri,shadows,shapes.geometric,shapes.misc,spy,trees,arrows.meta]
\newcommand{\im}{\mathrm i}
\newcommand{\del}{\partial}

\newcommand{\vkappa}{\varkappa}

\newcommand{\vtheta}{\vartheta}
\newcommand{\wt}{\widetilde}
\newcommand{\mc}{\mathcal}
\newcommand{\ms}{\mathsf}
\newcommand{\dd}{\mathrm{d}}

\newcommand{\ket}[1]{\left|#1\right\rangle}      
\newcommand{\bra}[1]{\left\langle #1\right|}     
\newcommand{\eq}{\begin{equation}}
\newcommand{\en}{\end{equation}}
\newcommand{\bear}{\begin{eqnarray}}
\newcommand{\ear}{\end{eqnarray}}

\usetikzlibrary{arrows,shapes,positioning}
\usetikzlibrary{decorations.markings}
\tikzstyle arrowstyle=[scale=1]
\tikzstyle directed=[postaction={decorate,decoration={markings,
		mark=at position .65 with {\arrow[arrowstyle]{stealth}}}}]
\tikzstyle reverse directed=[postaction={decorate,decoration={markings,
		mark=at position .65 with {\arrowreversed[arrowstyle]{stealth};}}}]

\title{Arctic curve of the free-fermion six-vertex model with reflecting end boundary condition}

\author{I.R. Passos and G.A.P. Ribeiro\footnote{E-mail: pavan@df.ufscar.br}}
\affil{Departamento de F\'{i}sica, Universidade Federal de S\~ao Carlos \\ S\~ao Carlos, SP 13565-905, Brazil}
\date{}
  
\begin{document}

\maketitle
\thispagestyle{empty}

\begin{abstract}
We consider the six-vertex model with reflecting end boundary condition. We study the asymptotic behavior of the boundary correlations. This asymptotic behavior is used as an input into the Tangent Method in order to derive 
analytically the arctic curve at the free fermion point. The obtained curve is a semicircle, which is in agreement with previous Monte Carlo simulations.
\end{abstract}

\newpage

\pagestyle{plain}
\pagenumbering{arabic}

\section{Introduction}

The six-vertex model is one of the most important integrable models and it has been largely studied over the years \cite{BAXTER,BOOK}. Its usefulness exceeds the classical statistical mechanics where it was first proposed and expresses itself through applications from mathematics and combinatorics \cite{KUPERBERG} to experimental developments involving artificial spin systems \cite{BRAMWELL,WANG}.
Besides, it also presents a number of special properties, e.g. the dependency of its physical properties on boundary conditions. 

This model was investigated under periodic, anti-periodic and also various instances of fixed boundary conditions \cite{LIEB,WU,OWCZAREK,BATCHELOR,KOREPIN2000,ZINNJUSTIN,BLEHER,RIBEIRO2015b,PARTIAL,GALLEAS,BLEHER2017}. 
The dependency on boundary conditions in the thermodynamic limit was firstly observed for the domain wall boundary \cite{KOREPIN1982,KOREPIN1992}, 
when it was shown that the thermodynamic properties in this case are different from the result for periodic boundary conditions \cite{KOREPIN2000,ZINNJUSTIN,BLEHER}. This fact suggested the existence of spatial phase separation, which was corroborated numerically \cite{NUM-DWBC}.

This spatial phase separation is due to the ice-rule, which restricts the number of configurations in the case of fixed boundary conditions \cite{KOREPIN2000,RIBEIRO2015b}. Even when the parameters of the system are adjusted for the disordered regime, certain fixed boundary conditions induce the formation of ordered regions of macroscopic size which propagate towards the bulk.
The separation lines between the ordered and disordered regions are the so-called arctic curves. For the six-vertex model with domain wall boundary condition, the analytical determination of the arctic curves in the disordered regime was carried out in \cite{ARTIC}. 

Besides the fact that arctic curves are important on their own right, they have also shown substantial role in quantum quenches, nonequilibrium transport in one-dimensional quantum spin chains and spin-ice models \cite{ARTICrole}.

More recently, the six-vertex model with reflecting end boundary condition \cite{TSUCHIYA} was also considered. In this case, the evaluation of thermodynamic properties \cite{RIBEIRO2015a} and the boundary correlations \cite{PASSOS} were analytically done. This was made possible thanks to the fact that the partition function of the six-vertex model with reflecting end boundary can be built from the Bethe state defined in Sklyanin construction for open spin chains \cite{SKYLIANIN}. In addition, the partition function in this case can also be represented as a determinant \cite{TSUCHIYA}.
Once again, it was observed that the free energy for the six-vertex model in this case differs from the one with periodic boundary conditions \cite{RIBEIRO2015a}, which motivated numerical investigation of spatial separation of phases \cite{LYBERG}.

Nevertheless, there is no analytical description of the arctic curves in the case of reflecting end boundary. In this paper, we present the analytical derivation of the arctic curve in a special case of free fermion point $\Delta=0$. The obtained curve is a semicircle, which is in good agreement
with the Monte Carlo simulation \cite{LYBERG}. In order to do that we have exploited the fact that the leading contribution for the free energy is determined by a solution of the Liouville equation \cite{RIBEIRO2015a}. This allowed us to obtain the asymptotic behavior of the boundary correlation on the free fermion point. We have also determined the point of contact of the curve with one of the boundaries. Finally, using the Tangent Method devised by Colomo and Sportiello in \cite{TANGENT}, we were able to analytically derive the arctic curve.

This paper is organized as follows. In section \ref{sixvertex}, we introduce the six-vertex model and the reflecting end boundary. In section \ref{bound-corr} we discuss the necessary boundary correlations and we study their asymptotic behavior. The contact point of the arctic curve with the boundary is obtained via one of these correlations. 
Finally, in section \ref{tangent} we use the results of previous sections in order to apply the Tangent Method to the determination of the arctic curve. Our conclusions are given in section \ref{conclusion}.

\section{The six-vertex model with reflecting end boundary condition}\label{sixvertex}

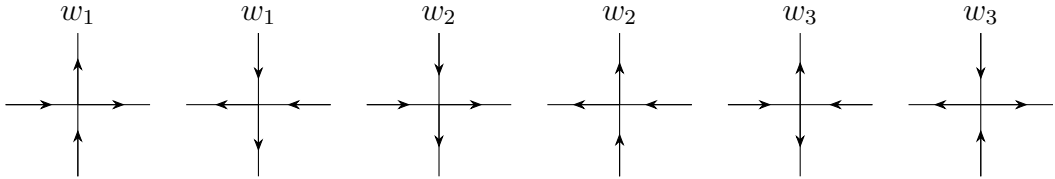
\begin{figure}[h!]
	\centering
	\begin{tikzpicture}[scale=.95,>=Stealth]
	
	\draw (-3,2.25) node {$w_1$};
	\draw (-3,0) -- (-3,2);
	\draw (-4,1) -- (-2,1);		
	\draw  [->] (-4,1) -- (-3.35,1);
	\draw  [->] (-3,1) -- (-2.35,1);
	\draw  [->] (-3,1) -- (-3,1.65);
	\draw  [->] (-3,0) -- (-3,.65);		
	
	\draw (-.5,2.25) node {$w_1$};			
	\draw (-.5,0) -- (-.5,2);
	\draw (-1.5,1) -- (.5,1);		
	\draw  [->] (.5,1) -- (-.1,1);
	\draw  [->] (-0.5,1) -- 	(-1.1,1);
	\draw  [->] (-.5,2) -- (-.5,1.35);
	\draw  [->] (-.5,1) -- (-.5,.35);

	\draw (2,2.25) node {$w_2$};			
	\draw (2,0) -- (2,2);
	\draw (1,1) -- (3,1);
	\draw [->] (1,1) -- (1.6,1);
	\draw [->] (2,1) -- (2.6,1);
	\draw [->] (2,2) -- (2,1.4);
	\draw [->] (2,1) -- (2,.4);
	
	\draw (4.5, 2.25) node {$w_2$};	
	\draw (4.5,0) -- (4.5,2);
	\draw (3.5,1) -- (5.5,1);
	\draw [->] (4.5,1) -- 	(3.85,1);			
	\draw [->] (5.5,1) -- 	(4.85,1);		
	\draw [->] (4.5,0) -- (4.5,0.6);
	\draw [->] (4.5,1) -- (4.5,1.6);	
	
	\draw (7, 2.25) node {$w_3$};			
	\draw (7,0) -- (7,2);
	\draw (6,1) -- (8,1);	
	\draw [->] (7,1) -- (7,1.6);
	\draw [->] (7,1) -- (7,.4);
	\draw [->] (6,1)-- (6.6,1);
	\draw [->] (8,1) -- (7.4,1);
	
	\draw (9.5, 2.25) node {$w_3$};		
	\draw (9.5,0) -- (9.5,2);
	\draw (8.5,1) -- (10.5,1);	
	\draw  [->] (9.5,1) -- (10.15,1);			
	\draw  [->] (9.5,1) -- 	(8.85,1);		
	\draw  [->] (9.5,2) -- (9.5,1.35);		
	\draw  [->] (9.5,0) -- (9.5,.65);			
	\end{tikzpicture}
	\caption{The Boltzmann weights of the six-vertex model.}
	\label{6v}
\end{figure}

The six-vertex model is an important integrable model of classical statistical mechanics \cite{BAXTER,BOOK}. It is defined on a rectangular lattice with arrows assigned to its edges, whose orientations are selected according to the ice-rule: two arrows point into and two arrows point away from each lattice vertex. This is depicted in Figure \ref{6v}, where we assigned to each vertex a Boltzmann weight $w_i$ invariant under arrow-reversal. 

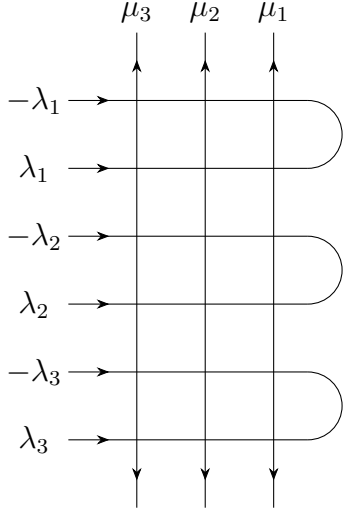
\begin{figure}[h!]
	\centering
	\begin{tikzpicture}[>=Stealth,scale=.9]
		
		\foreach \x in {0,...,2}
		{  
			\draw [<->] (\x,0.4) -- (\x,6.6);
			\draw (\x,0) -- (\x,0.5);
			\draw (\x,6.5) -- (\x,7);
	   	}
   	
   		\foreach \x in {1,...,6}
   		{	
   			\draw (-.5,\x) -- (2.5,\x);
   			\draw [->] (-1,\x) -- (-.4,\x);	
   		}
   	
   		\foreach \x in {1,3,5}
   		{		
   			\draw (2.5,\x) arc [start angle=-90,end angle=90, radius=.5cm];
   		}
   	
   		\draw (0,7.3) node {$\mu_3$};
   		\draw (1,7.3) node {$\mu_2$};
   		\draw (2,7.3) node {$\mu_1$};
   		
   		\draw (-1.5,1) node {$\lambda_3$};
   		\draw (-1.5,2) node {$-\lambda_3$};
   		\draw (-1.5,3) node {$\lambda_2$};
   		\draw (-1.5,4) node {$-\lambda_2$};
   		\draw (-1.5,5) node {$\lambda_1$};
   		\draw (-1.5,6) node {$-\lambda_1$};
	\end{tikzpicture}
	\caption{Partition function of the six-vertex model with reflecting end boundary condition for $N=3$.}
	\label{latt-rdwbc}
\end{figure}

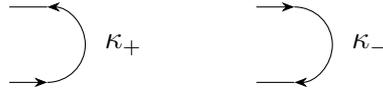
\begin{figure}[h!]
	\centering
	\begin{tikzpicture}[scale=.5,>=Stealth]
		
		\draw [->] (0,0) -- (1,0);
		\draw (0,2) -- (1,2);
		\draw [->] (1,0) arc [start angle=-90, end angle=90, radius,radius=1cm];
		\draw (3,1) node {$\kappa_+$};

		\draw (6.5,0) -- (7.5,0);
		\draw [->] (6.5,2) -- (7.5,2);
		\draw [<-] (7.5,0) arc [start angle=-90, end angle=90, radius,radius=1cm];
		\draw (9.5,1) node {$\kappa_-$};
	\end{tikzpicture}
	\caption{Elements of the $K$-matrix.}
	\label{v-refl}
\end{figure}

We consider the six-vertex model in a $2N\times N$ rectangular lattice with domain wall boundary conditions and one reflecting end, as illustrated in Figure \ref{latt-rdwbc}. 
We assume the only possible configurations for the reflecting boundary vertices are those represented in Figure \ref{v-refl}. 

In the case of an inhomogeneous model, the Boltzmann weights are site-dependent, which is expressed through two sets of parameters, namely $\{\lambda_j\}_{j=1}^{N}$ for the double rows and $\{\mu_k\}_{k=1}^{N}$ for the columns. Then, the weights of vertices on even rows (counting from the top) are given by
\begin{align}
	w_1=a_-(\lambda_j,\mu_k),&&w_2=b_-(\lambda_j,\mu_k),&&w_3=c_-(\lambda_j,\mu_k),
	\label{wminus}
\end{align}
with $j,k=1,\ldots,N$, whereas the weights of odd rows are
\begin{align}
	w_1=b_+(\lambda_j,\mu_k),&&w_2=a_+(\lambda_j,\mu_k),&&w_3=c_+(\lambda_j,\mu_k),
	\label{wplus}
\end{align}
as a consequence of the reflection on the left boundary (which reverses the sign of horizontal spectral parameters $\lambda\rightarrow-\lambda$). 
Throughout this paper, we denote $v_{\pm}(\lambda,\mu)=v(\lambda\pm\mu)$ where $v\in\{a,b,c\}$.

For the six-vertex model, the Boltzmann weights \eqref{wminus}, \eqref{wplus} can be seen as entries of the $R$-matrix,
\begin{align}
R(\lambda)=\begin{pmatrix}
a(\lambda) & 0 & 0 & 0 \\
0 & b(\lambda) & c(\lambda) & 0 \\
0 & c(\lambda) & b(\lambda) & 0 \\
0 & 0 & 0 & a(\lambda)
\end{pmatrix},
\label{rmatrix}
\end{align}
which is a solution of the Yang-Baxter equation,
\begin{align}
R_{12}(\lambda-\mu)R_{13}(\lambda)R_{23}(\mu)=R_{23}(\mu)R_{13}(\lambda)R_{12}(\lambda-\mu).
\label{yb}
\end{align}
As a consequence of \eqref{yb}, the weights $a(\lambda)$, $b(\lambda)$ and $c(\lambda)$ must satisfy (for all $\lambda$)
\begin{align}
	\Delta=\frac{a^2+b^2-c^2}{2ab},
	\label{inv}
\end{align}
where $\Delta$ is a constant. For $|\Delta|<1$ (disordered regime), we choose the parametrization
\begin{align}
	a(\lambda)=\sin(\lambda+2\eta),&& b(\lambda)=\sin(\lambda),&&
	c(\lambda)=\sin(2\eta),
	\label{par-d}
\end{align}
and thus $\Delta=\cos(2\eta)$. Since $c(\lambda)$ is a constant, we omit its argument from now on. 

On the other hand, the boundary weights $\kappa_+(\lambda)$, $\kappa_-(\lambda)$ are matrix elements of the $K$-matrix,
\begin{align}
	K(\lambda)=\begin{pmatrix}
	\kappa_+(\lambda) & 0 \\
	0 & \kappa_-(\lambda)
	\end{pmatrix},
\end{align}
which must satisfy the reflection equation
\begin{align}
	R_{12}(\lambda-\mu)K_1(\lambda)R_{12}(\lambda+\mu)K_2(\mu)=K_2(\mu)R_{12}(\lambda+\mu)K_1(\lambda)R_{12}(\lambda-\mu),
	\label{refl-w}
\end{align}
in order to preserve integrability on the boundary \cite{SKYLIANIN}. For $|\Delta|<1$, the solution of \eqref{refl-w} reads
\begin{align}
	\kappa_{\pm}(\lambda)=\frac{\sin(\xi\pm\lambda)}{\sin(\xi)},
\end{align}
where $\xi$ is the boundary parameter.

Let $\mc{H}_l$, $\mc{V}_k$ be two-dimensional vector spaces associated to the $l$th row (from the top) and $k$th column (from the right), with $l=1,\ldots,2N$ and $k=1,\ldots,N$. The horizontal (vertical) space of all rows (columns) is the tensor product $\mc{H}=\otimes_{l=1}^{2N}\mc{H}_l$ $\left(\mc{V}=\otimes_{k=1}^{N}\mc{V}_k\right)$. Thus, the $R$-matrix associated to the vertex at ($l,k$) position acts non trivially only on the vector space $\mc{H}_l\otimes \mc{V}_k$. The ordered product of $R$-matrices along an even row $l=2j$ gives rise to the monodromy matrix $\mc{T}(\lambda_j)$, which can be written as a $2\times 2$ matrix on the local horizontal space $\mc{H}_{2j}$,
\begin{align}
	\mc{T}(\lambda_j)=R_{jN}(\lambda_j-\mu_N)\ldots R_{j1}(\lambda_j-\mu_1)=\begin{pmatrix}
	A(\lambda_j) & B(\lambda_j) \\
	C(\lambda_j) & D(\lambda_j)
	\end{pmatrix},
\end{align}
with $A,B,C,D$ operators acting on the vertical space $\mc{V}$. Similarly, on the odd rows $l=2j-1$ we have
\begin{align}
	\widetilde{\mc{T}}(\lambda_j)=R_{j1}(\lambda_j+\mu_1)\ldots R_{jN}(\lambda_j+\mu_N)=\begin{pmatrix}
	\widetilde{A}(\lambda_j) & \widetilde{B}(\lambda_j) \\
	\widetilde{C}(\lambda_j) & \widetilde{D}(\lambda_j)
	\end{pmatrix},
\end{align}
acting on the $\mc{H}_{2j-1}\otimes\mc{V}$ space. In terms of $\mc{T}(\lambda)$ and $\wt{\mc{T}}(\lambda)$, we introduce the Sklyanin's monodromy matrix $\mc{U}(\lambda)$,
\begin{align}
	\mc{U}(\lambda)=\mc{T}(\lambda)K(\lambda)\wt{\mc{T}}(\lambda)=\begin{pmatrix}
	\mc{A}(\lambda) & \mc{B}(\lambda) \\
	\mc{C}(\lambda) & \mc{D}(\lambda)
	\end{pmatrix},
\end{align}
which satisfies the reflection algebra 
\begin{align}
R_{12}(\lambda-\mu)\mc{U}_1(\lambda)R_{12}(\lambda+\mu)\mc{U}_2(\mu)=\mc{U}_2(\mu)R_{12}(\lambda+\mu)\mc{U}_1(\lambda)R_{12}(\lambda-\mu).
\label{refl-mon}
\end{align}

The partition function of the six-vertex model with reflecting end boundary conditions in a $2N\times N$ lattice (depicted in Figure \ref{latt-rdwbc}) can be defined in terms of the monodromy matrix element $\mc{B}(\lambda)$ as follows
\begin{align}
	Z_N(\{\lambda_j\},\{\mu_k\})=\bra{\Downarrow}\mc{B}(\lambda_1)\ldots\mc{B}(\lambda_N)\ket{\Uparrow},
	\label{partition}
\end{align}
where $\ket{\Uparrow}=\ket{\uparrow\ldots\uparrow}$, $\ket{\Downarrow}=\ket{\downarrow\ldots\downarrow}$ are the up and down ferromagnetic states of the vertical space $\mc{V}$. The partition function \eqref{partition} admits a determinant representation, due to Tsuchiya \cite{TSUCHIYA}, which is given by
\begin{align}
	Z_N(\{\lambda_j\},\{\mu_k\})&=\frac{\prod_{j,k=1}^{N}a_+(\lambda_j,\mu_k) a_-(\lambda_j,\mu_k) b_+(\lambda_j,\mu_k) b_-(\lambda_j,\mu_k)}{\prod_{k<j}^{N}a_+(\lambda_j,\lambda_k)b_-(\lambda_j,\lambda_k)\prod_{m<n}^{N}b_+(\mu_m,\mu_n)b_-(\mu_m,\mu_n)}\nonumber \\
	&\times\prod_{j=1}^{N}b(2\lambda_j)\kappa_-(\mu_j)\det\ms{M},
	\label{part-tsu}
\end{align}
where the entries of $\ms{M}$ are
\begin{align}
	\ms{M}_{jk}=\psi(\lambda_j,\mu_k),&&j,k=1,\ldots,N,&&\psi(\lambda,\mu)=\frac{c}{a_+(\lambda,\mu) a_-(\lambda,\mu) b_+(\lambda,\mu) b_-(\lambda,\mu)}.
\end{align}

In order to study the model in the thermodynamic limit, we must first take the homogeneous limit $\lambda_1,\ldots,\lambda_N\rightarrow \lambda$, $\mu_1,\ldots,\mu_N\rightarrow \mu$ of the partition function \eqref{part-tsu}. This is done along the lines of \cite{KOREPIN1992}. The final result reads
\begin{align}
	Z_N(\lambda,\mu)=\frac{[a_+(\lambda,\mu) a_-(\lambda,\mu) b_+(\lambda,\mu) b_-(\lambda,\mu)]^{N^2}}{C_N[-a(2\lambda)b(2\mu)]^{N(N-1)/2}}[b(2\lambda)\kappa_-(\mu)]^{N}\tau_N(\lambda,\mu),
	\label{part-hom}
\end{align}
where $C_N=\left[\prod_{j=0}^{N-1}j!\right]^2$ and
\begin{align}\tau_N(\lambda,\mu)=\det\overline{\ms{M}},&&\overline{\ms{M}}_{jk}=\del_{\lambda}^{j-1}\del_{\mu}^{k-1}\psi(\lambda,\mu),&& j,k=1,\ldots,N.
\label{taufunc}
\end{align}

From \eqref{part-hom} we can find all thermodynamic quantities of the model, such as free energy. Indeed, it was shown \cite{RIBEIRO2015a} the free energy per vertex $F(\lambda,\mu)$ in the disordered regime is given by
\begin{align}
	e^{-2F(\lambda,\mu)}=\frac{a_+(\lambda,\mu) a_-(\lambda,\mu) b_+(\lambda,\mu) b_-(\lambda,\mu)}{[-a(2\lambda)b(2\mu)]^{1/2}}e^{2 f(\lambda,\mu)},
	\label{free-en}
\end{align}
where $f(\lambda,\mu)$ satisfies the Liouville equation
\begin{align}
	2\del_{\lambda}\del_{\mu}f(\lambda,\mu)=e^{4f(\lambda,\mu)},
\end{align}
whose solution is fixed, thanks to boundary conditions on the partition function \eqref{part-hom}, as
\begin{align}
	e^{4f(\lambda,\mu)}=-\frac{\alpha^2 \sin(\alpha\lambda)\sin(\alpha\mu)}{[\cos(\alpha\mu)-\cos(\alpha\lambda)]^2},&&\alpha=\frac{\pi}{\eta}.
	\label{liou}
\end{align}
It is worth to recall the free energy \eqref{free-en} for the six-vertex model with reflecting end boundary condition differs from the result in the case of periodic boundary. 
This discrepancy can be understood in terms of the number of configurations: because of the fixed boundary, the admissible states in the former are severely restricted in comparison to the latter.

In the scaling limit, the effect of the domain wall boundary with a reflecting end is manifested through phase separation phenomena, in which ferroelectric and disordered regions coexist and are delimited by the arctic curves. In the following sections, we determine analytically the arctic curve in the special point $\Delta=0$, $\mu=0$, $a=b$. To achieve this, we first discuss two types of boundary correlation functions for this model.

\section{Boundary correlations}\label{bound-corr}

In this section, we review two types of boundary correlations necessary to the determination of the arctic curve and its contact points. For the six-vertex model with reflecting end boundary condition, these correlations were first introduced in \cite{PASSOS}. 

The first correlation function, $G_N^{(r)}$, describes the probability that the polarization state of boundary vertical edge between the $r$th and $(r+1)$th double rows is a down arrow. We can define it as 
\begin{align}
	G_N^{(r)}=\frac{1}{Z_N}\bra{\Downarrow}\mc{B}(\lambda_N)\ldots\mc{B}(\lambda_{r+1})q_N\mc{B}(\lambda_r)\ldots\mc{B}(\lambda_1)\ket{\Uparrow}.
	\label{corr-g}
\end{align}
Meanwhile, the second correlation, $H_N^{(r)}$, reflects the fact that there must a sole
$c$-vertex in the first column (from the left) for this boundary condition. The probability that this vertex is placed at either of the stripes of the $r$th double row can be written as
\begin{align}
	H_N^{(r)}=\frac{1}{Z_N}\bra{\Downarrow}\mc{B}(\lambda_N)\ldots\mc{B}(\lambda_{r+1})q_N\mc{B}(\lambda_r)p_N\mc{B}(\lambda_{r-1})\ldots\mc{B}(\lambda_1)\ket{\Uparrow}.
	\label{corr-h}
\end{align}
The operators $q_N=\frac{1}{2}(1-\sigma_N^z)$, $p_N=\frac{1}{2}(1+\sigma_N^z)$ are projectors onto spin-down and spin-up states, respectively. From \eqref{corr-g} and \eqref{corr-h} it is clear that these functions are related by
\begin{align}
	G_N^{(r)}=\sum_{j=1}^{r}H_N^{(j)}.
	\label{rec}
\end{align}
Moreover, from \eqref{partition} we see that $G_N^{(N)}=1$. Using the reflection algebra one can find recurrence relations between the correlations $G_N^{(r)}$ and $H_N^{(r)}$ and the partition function of $2(N-1)\times (N-1)$ rectangular sublattices \cite{PASSOS}. Then, using Tsuchiya formula \eqref{part-tsu}, these correlations can also be written in terms of determinants of $N\times N$ matrices. 

Another important quantity regarding these correlations is the generating function $h_N(z)$, formally defined as
\begin{align}
	h_N(z)=\sum_{r=1}^{N}H_N^{(r)}z^{r-1},&&z\in\mathbb{C}.
	\label{gen}
\end{align}
Note that $h_N(1)=1$ thanks to \eqref{rec}. From Cauchy integral formula, we can invert this relation in order to obtain $H_N^{(r)}$ as
\begin{align}
	H_N^{(r)}=\frac{1}{2\pi\im}\oint_C \frac{h_N(z)}{z^r}\dd z,
	\label{int-h}
\end{align}
where $C$ is a counterclockwise oriented closed path around the origin. From \eqref{rec} and \eqref{int-h} we also obtain an integral formula for $G_N^{(r)}$, which reads
\begin{align}
	G_N^{(r)}=-\frac{1}{2\pi\im}\oint_C\frac{h_N(z)}{(z-1)z^r} \dd z.
	\label{int-g}
\end{align}
As we shall see below, the integral formula \eqref{int-g} allows one to obtain the contact point between arctic curve and the left boundary, while the function $H_N^{(r)}$ plays a fundamental role in the derivation of an analytical expression for the curve through its generating function $h_N(z)$. In order to achieve both goals, we need the asymptotic behavior of $h_N(z)$ in the scaling limit.

\subsection{Asymptotic behavior of $h_N(z)$}
Here, we will determine how the function $h_N(z)$ behaves in the large $N$ limit. In order to do this, we first establish a connection between this function and the partition function of a partially 
inhomogenenous model. 

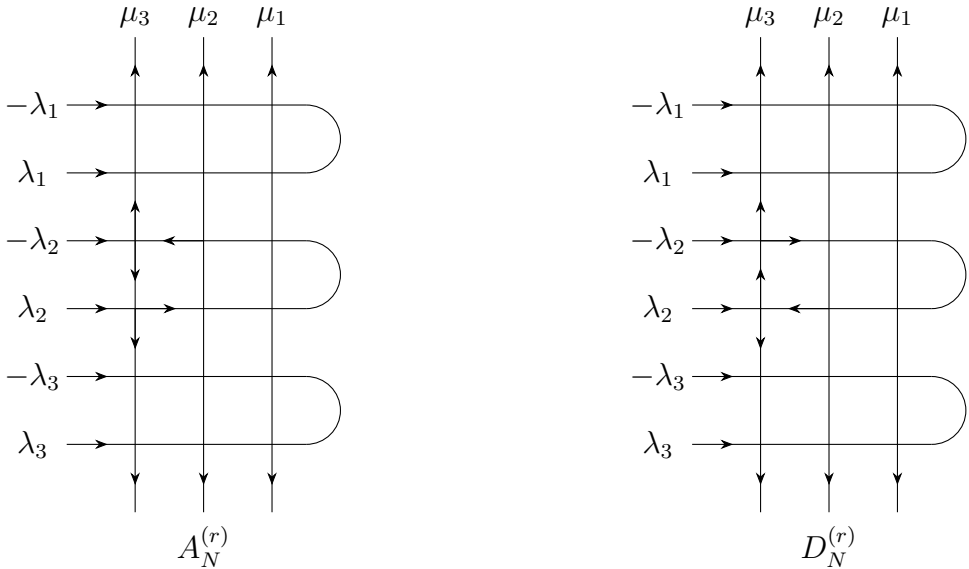
\begin{figure}[h!]
	\centering
	\begin{minipage}{.4\textwidth}
		\begin{tikzpicture}[>=Stealth,scale=.9]
		
		\foreach \x in {0,...,2}
		{  
			\draw [<->] (\x,0.4) -- (\x,6.6);
			\draw (\x,0) -- (\x,0.5);
			\draw (\x,6.5) -- (\x,7);
		}
		
		\foreach \x in {1,...,6}
		{	
			\draw (-.5,\x) -- (2.5,\x);
			\draw [->] (-1,\x) -- (-.4,\x);	
		}
		
		\foreach \x in {1,3,5}
		{		
			\draw (2.5,\x) arc [start angle=-90,end angle=90, radius=.5cm];
		}
		
		\draw [->] (0,4) -- (0,4.6);
		\draw [->] (1,4) -- (.4,4);
		\draw [->] (0,4) -- (0,3.4);
		\draw [->] (0,3) -- (0,2.4);
		\draw [->] (0,3) -- (0.6,3);
		
		\draw (0,7.3) node {$\mu_3$};
		\draw (1,7.3) node {$\mu_2$};
		\draw (2,7.3) node {$\mu_1$};
		
		\draw (-1.5,1) node {$\lambda_3$};
		\draw (-1.5,2) node {$-\lambda_3$};
		\draw (-1.5,3) node {$\lambda_2$};
		\draw (-1.5,4) node {$-\lambda_2$};
		\draw (-1.5,5) node {$\lambda_1$};
		\draw (-1.5,6) node {$-\lambda_1$};
		
		\draw (1,-.5) node {$A_N^{(r)}$};
		\end{tikzpicture}
	\end{minipage}
	\hfill
		\begin{minipage}{.4\textwidth}
		\begin{tikzpicture}[>=Stealth,scale=.9]
		
		\foreach \x in {0,...,2}
		{  
			\draw [<->] (\x,0.4) -- (\x,6.6);
			\draw (\x,0) -- (\x,0.5);
			\draw (\x,6.5) -- (\x,7);
		}
		
		\foreach \x in {1,...,6}
		{	
			\draw (-.5,\x) -- (2.5,\x);
			\draw [->] (-1,\x) -- (-.4,\x);	
		}
		
		\foreach \x in {1,3,5}
		{		
			\draw (2.5,\x) arc [start angle=-90,end angle=90, radius=.5cm];
		}
		
		\draw [->] (0,4) -- (0,4.6);
		\draw [->] (0,4) -- (0.6,4);
		\draw [->] (0,3) -- (0,3.6);
		\draw [->] (0,3) -- (0,2.4);
		\draw [->] (1,3) -- (0.4,3);
		
		\draw (0,7.3) node {$\mu_3$};
		\draw (1,7.3) node {$\mu_2$};
		\draw (2,7.3) node {$\mu_1$};
		
		\draw (-1.5,1) node {$\lambda_3$};
		\draw (-1.5,2) node {$-\lambda_3$};
		\draw (-1.5,3) node {$\lambda_2$};
		\draw (-1.5,4) node {$-\lambda_2$};
		\draw (-1.5,5) node {$\lambda_1$};
		\draw (-1.5,6) node {$-\lambda_1$};
		\draw (1,-.5) node {$D_N^{(r)}$};
		\end{tikzpicture}
	\end{minipage}
	\caption{The functions $A_N^{(r)}$ and $D_N^{(r)}$ for $N=3$ and $r=2$.}
	\label{a-d}
\end{figure}

We start by noticing that $H_N^{(r)}$ \eqref{corr-h} can be seen as the sum of two terms,
\begin{align}
	H_N^{(r)}=\frac{A_N^{(r)}+D_N^{(r)}}{Z_N},
\end{align}
depending on whether the $c$-vertex is placed at the top or bottom horizontal line of the $r$th double row. Graphical representations of $A_N^{(r)}$ and $D_N^{(r)}$ are depicted on Figure \ref{a-d}.

Now consider a partially inhomogeneous model, with spectral parameters
\begin{align}
	\lambda_1=\ldots=\lambda_N=\lambda,&&\mu_1=\ldots=\mu_{N-1}=\mu,&&\mu_N=\mu+\omega.
\end{align}
From Figure \ref{a-d} we can see that
\begin{align}
	A_N^{(r)}(\lambda,\mu,\omega)&=\left[\frac{a_+(\lambda,\mu+\omega)b_-(\lambda,\mu+\omega)}{a_+(\lambda,\mu)b_-(\lambda,\mu)}\right]^{N-r}\left[\frac{a_-(\lambda,\mu+\omega)b_+(\lambda,\mu+\omega)}{a_-(\lambda,\mu)b_+(\lambda,\mu)}\right]^{r-1}\times \nonumber \\
	&\frac{b_-(\lambda,\mu+\omega)}{b_-(\lambda,\mu)}A_N^{(r)}(\lambda,\mu),\\
	D_N^{(r)}(\lambda,\mu,\omega)&=\left[\frac{a_+(\lambda,\mu+\omega)b_-(\lambda,\mu+\omega)}{a_+(\lambda,\mu)b_-(\lambda,\mu)}\right]^{N-r}\left[\frac{a_-(\lambda,\mu+\omega)b_+(\lambda,\mu+\omega)}{a_-(\lambda,\mu)b_+(\lambda,\mu)}\right]^{r-1}\times\nonumber \\
	&\frac{b_+(\lambda,\mu+\omega)}{b_+(\lambda,\mu)}D_N^{(r)}(\lambda,\mu),
\end{align}
where $A_N^{(r)}(\lambda,\mu)$ and $D_N^{(r)}(\lambda,\mu)$ compose the function $H_N^{(r)}(\lambda,\mu)$ for the fully homogeneous model ($\omega=0$). 
Since $\sum_{r=1}^{N}H_N^{(r)}=1$, we can relate the partially inhomogeneous and homogeneous models as
\begin{align}
	Z_N(\lambda,\mu,\omega)&=\sum_{r=1}^{N}\left[\frac{b_-(\lambda,\mu+\omega)}{b_-(\lambda,\mu)}A_N^{(r)}(\lambda,\mu)+\frac{b_+(\lambda,\mu+\omega)}{b_+(\lambda,\mu)}D_N^{(r)}(\lambda,\mu)\right]\times\nonumber \\
	&\left[\frac{a_+(\lambda,\mu+\omega)b_-(\lambda,\mu+\omega)}{a_+(\lambda,\mu)b_-(\lambda,\mu)}\right]^{N-r}\left[\frac{a_-(\lambda,\mu+\omega)b_+(\lambda,\mu+\omega)}{a_-(\lambda,\mu)b_+(\lambda,\mu)}\right]^{r-1}.
\end{align}
In the special case when $\mu=0$ and $\Delta=0\  (\eta=\pi/4)$, we can drop the $\pm$ indices and the parametrization \eqref{par-d} becomes
\begin{align}
	a(\lambda)=\cos(\lambda),&&b(\lambda)=\sin(\lambda),&&c(\lambda)=1.
	\label{par-delta0}
\end{align}
Then we can combine $Z_N(\lambda,\omega)$ and $Z_N(\lambda,\pi/2-\omega)$ to obtain
\begin{align}
	\cos(\omega)Z_N(\lambda,\omega)-&\sin(\omega)Z_N(\lambda,\pi/2-\omega)=\cos(2\omega)Z_N(\lambda)\times \nonumber\\
	&\sum_{r=1}^{N}H_N^{(r)}(\lambda)\left[\frac{a(\lambda+\omega)b(\lambda-\omega)}{a(\lambda)b(\lambda)}\right]^{N-r}\left[\frac{a(\lambda-\omega)b(\lambda+\omega)}{a(\lambda)b(\lambda)}\right]^{r-1}.
	\label{z-h1}	
\end{align}
Using the definition \eqref{gen}, relation \eqref{z-h1} can be rewritten as
\begin{align}
	\frac{a(\omega)Z_N(\lambda,\omega)-b(\omega)Z_N(\lambda,\pi/2-\omega)}{a(2\omega)Z_N(\lambda)}=\left[\frac{a(\lambda+\omega)b(\lambda-\omega)}{a(\lambda)b(\lambda)}\right]^{N-1}h_N[\gamma(\omega)],
	\label{rat-gen}
\end{align}
where
\begin{align}
	\gamma(\omega)=\frac{a(\lambda-\omega) b(\lambda+\omega)}{a(\lambda+\omega) b(\lambda-\omega)}.
	\label{gam}
\end{align}

From \eqref{rat-gen} we see that the asymptotic behavior of $h_N(z)$ is determined once we know the ratio $Z_N(\lambda,\omega)/Z_N(\lambda)$ in the large $N$ limit. Using \eqref{part-tsu} and \eqref{part-hom}, in the general case $\mu\ne0$ we have
\begin{align}
	\frac{Z_N(\lambda,\mu,\omega)}{Z_N(\lambda,\mu)}&=\frac{(N-1)!}{[b(\omega)]^{N-1}}\frac{\kappa_-(\mu+\omega)}{\kappa_-(\mu)}\left[\frac{b(2\mu)}{b(2\mu+\omega)}\right]^{N-1}\times\nonumber \\
	&\left[\frac{a_+(\lambda,\mu+\omega) a_-(\lambda,\mu+\omega) b_+(\lambda,\mu+\omega) b_-(\lambda,\mu+\omega)}{a_+(\lambda,\mu) a_-(\lambda,\mu) b_+(\lambda,\mu) b_-(\lambda,\mu)}\right]^N S_N(\mu,\omega),
	\label{ratio-part}
\end{align}
where
\begin{align}
	S_N(\mu,\omega)=\frac{\widetilde{\tau}_N(\lambda,\mu,\omega)}{\tau_N(\lambda,\mu)},
	\label{ratio}
\end{align}
with
\begin{align}	
	\widetilde{\tau}_N=\det\wt{\ms{M}},&&\wt{\ms{M}}_{jk}=\begin{cases}
	\del_{\lambda}^{j-1}\del_{\mu}^{k-1}\psi(\lambda,\mu), & k\ne N \\
	\del_{\lambda}^{j-1}\psi(\lambda,\mu+\omega), & k=N
	\end{cases} &&j,k=1,\ldots,N.
\end{align}
Note $\lambda$ is seen as a parameter in $S_N(\mu,\omega)$, not a variable. In order to determine the asymptotic behavior of $S_N(\mu,\omega)$ we seek a differential equation solved by this ratio of determinants, similar to what have been done in \cite{ARTIC}, \cite{RIBEIRO2015a}. 

To this end, consider the Sylvester identity \cite{GANTMACHER}
\begin{align}
	\det\ms{A}=(\det\ms{A}_{[p+1,\ldots,n;p+1,\ldots,n]})^{-(n-p-1)}\det\ms{B},&& p\in[1,n],
\end{align}
where $\ms{A}$ is a $n\times n$ matrix and $\det\ms{A}_{[j;k]}$ denotes the determinant of a matrix obtained from $\ms{A}$ by excluding its $j$th row and $k$th column. The entries of $\ms{B}$ are minors of $\ms{A}$, given as
\begin{align}
	\ms{B}_{jk}=\det\ms{A}_{[p+1,\ldots,p+j-1,p+j+1,\ldots,n;p+1,\ldots,p+k-1,p+k+1,\ldots,n]}.
\end{align}
Taking $n=N+1$, $p=N-1$ and setting $\det\ms{A}=\tau_{N+1}$ from \eqref{taufunc}, we have
\begin{align}
	\tau_{N+1}=\frac{1}{\tau_{N-1}}\begin{vmatrix}
	\del_{\lambda}\del_{\mu}\tau_N & \del_{\lambda}\tau_N \\
	\del_{\mu}\tau_N & \tau_N 
	\end{vmatrix}.
	\label{bi-toda}
\end{align}
On the other hand, if $\det\ms{A}=\widetilde{\tau}_{N+1}$,
\begin{align}
\widetilde{\tau}_{N+1}=\frac{1}{\tau_{N-1}}\begin{vmatrix}
\del_{\lambda}\widetilde{\tau}_N &  \del_{\lambda}\tau_N \\
\widetilde{\tau}_N & \tau_N
\end{vmatrix}.
\label{bi-toda-t}
\end{align}
Moreover, by differentiating $\wt{\tau}_{N+1}$ with respect to $\mu$ and $\omega$, we find the relation
\begin{align}
	\del_{\mu}\wt{\tau}_{N+1}=\del_{\omega}\wt{\tau}_{N+1}+\frac{1}{\tau_{N-1}}
	\begin{vmatrix}
	\del_{\lambda}\wt{\tau}_N & \del_{\lambda}\del_{\mu}\tau_N \\
	\wt{\tau}_N& \del_{\mu}\tau_N
	\end{vmatrix}.
	\label{mu-om}
\end{align}
Using equations \eqref{bi-toda} and \eqref{bi-toda-t} in order to eliminate the derivatives in $\lambda$, as well as the definition \eqref{ratio}, we can bring \eqref{mu-om} into the form
\begin{align}
	\del_{\mu}S_N=\del_{\omega}S_N+S_N\del_{\mu}\left[\log\left(\frac{\tau_{N-1}}{\tau_N}\right)\right]-S_{N-1}.
	\label{diff}
\end{align}

In order to propose a formula for $S_N(\mu,\omega)$, first notice this function satisfies the boundary condition
\begin{align}
S_N(\mu,\omega)\sim\frac{\omega^{N-1}}{(N-1)!},&& \omega\rightarrow 0,
\label{bound}
\end{align}
since $Z_N(\lambda,\mu,\omega)/Z_N(\lambda,\mu)\rightarrow 1$ as $\omega\rightarrow 0$ (see equation \eqref{ratio-part}). On the other hand, $\tau_N$ has an exponential behavior \cite{RIBEIRO2015a}
\begin{align}
	\tau_N(\lambda,\mu)=C_N e^{2N^2 f(\lambda,\mu)+O(N)},
	\label{tau}
\end{align}
with $f(\lambda,\mu)$ given by \eqref{liou}. Since the determinants $\tau_N$ and $\wt{\tau}_N$ differ only by one column, we expect the behavior of $\wt{\tau}_N$ to be similar to \eqref{tau}. Therefore their ratio $S_N$ must increase exponentially as
\begin{align}
	S_N(\mu,\omega)=\frac{1}{(N-1)!}e^{N\Omega(\mu,\omega)+o(N)},
	\label{ansatz}
\end{align}
where $o(N)$ indicates terms whose order are smaller than $N$. Substitution of \eqref{ansatz} into \eqref{diff} leads to
\begin{align}
	(\del_{\mu}-\del_{\omega})\Omega(\mu,\omega)+4\del_{\mu}f(\lambda,\mu)+e^{-\Omega(\mu,\omega)}=0,
\end{align}
or, in terms of $W(\mu,\omega)=e^{\Omega(\mu,\omega)}$, this differential equation becomes
\begin{align}
	\left[\del_{\mu}-\del_{\omega}+4\del_{\mu}f(\lambda,\mu)\right]W(\mu,\omega)=-1,
\end{align}
whose solution is given by
\begin{align}
	W(\mu,\omega)=2e^{-4f(\lambda,\mu)}\del_{\lambda}[f(\lambda,\mu+\omega)-f(\lambda,\mu)].
	\label{ww}
\end{align}
Finally, after substituting \eqref{ww} into \eqref{ratio-part} and taking $\mu=0$, $\eta=\pi/4$, it follows
\begin{align}
	\frac{Z_N(\lambda,\omega)}{Z_N(\lambda)}=\frac{\kappa_-(\omega)}{[b(\omega)]^{2N}}\left[\frac{a(\lambda+\omega)a(\lambda-\omega)b(\lambda+\omega)b(\lambda-\omega)}{a^2(\lambda)b^2(\lambda)}\right]^N[g(\lambda,\omega)]^N,
	\label{ratio-part-f}
\end{align}
for large $N$, where
\begin{align}
	g(\lambda,\omega)=\frac{1}{4}\frac{b^2(2\lambda)b^2(2\omega)}{b(2\lambda-2\omega)b(2\lambda+2\omega)}.
\end{align}
For simplicity, we take the boundary parameter as $\xi=\pi/2$ so $\kappa_{\pm}(\lambda)=\cos(\lambda)$. Thus, after substituting \eqref{ratio-part-f} into \eqref{rat-gen} and solving for $h_N[\gamma(\omega)]$ we arrive at
\begin{align}
	h_N[\gamma(\omega)]=\frac{1}{a(2\omega)}\left(\frac{a^2(\omega)}{[b(\omega)]^{2N}}-\frac{b^2(\omega)}{[a(\omega)]^{2N}}\right)\left[\frac{a(\lambda-\omega)b(\lambda+\omega)}{a(\lambda)b(\lambda)}g(\lambda,\omega)\right]^N.
\end{align}
Notice that in order to the Boltzmann weights $a(\lambda\pm\omega)$, $b(\lambda\pm\omega)$ be real and positive in the parametrization \eqref{par-delta0}, the parameters $\lambda$ and $\omega$ are restricted as
\begin{align}
	0<\lambda\le \frac{\pi}{4},&&-\lambda<\omega<\lambda.
\end{align}
Therefore $|\tan(\omega)|<1$ and consequently $(b(\omega)/a(\omega))^{2(N+1)}\ll 1$ as $N\gg 1$. Hence
\begin{align}
	h_N[\gamma(\omega)]\sim\frac{a^2(\omega)}{a(2\omega)}\left[\frac{a(\lambda-\omega)b(\lambda+\omega)g(\lambda,\omega)}{a(\lambda)b(\lambda)b^2(\omega)}\right]^N.
	\label{asymp-h}
\end{align}
After taking the logarithm of \eqref{asymp-h} and its thermodynamic limit, we get the final expression
\begin{align}
	\lim_{N\rightarrow\infty}\frac{\log\ h_N[\gamma(\omega)]}{N}=\log\left[\frac{a^2(\omega)a(\lambda)b(\lambda)}{a(\lambda+\omega)b(\lambda-\omega)}\right].
	\label{asymp-log-h}
\end{align}

\subsection{Contact point}
Before tackling the problem of finding the analytical expression for the arctic curve, we now address the question of obtaining the location of contact point with the left boundary, following \cite{ARTIC}.

\begin{figure}[h!]
	\centering
	\begin{tikzpicture}
		\draw[color=black!50,dashed] (0,0) rectangle (2.5,5);
		\draw (0,-.3) node {\footnotesize$0$};
		\draw (2.5,-.3) node {\footnotesize$1$};
		\draw (-.3,5) node {\footnotesize$2$};
		\draw (-.3,2.5) node {\footnotesize$\vkappa$};
		\draw [->] (0,0) -- (3.5,0);
		\draw (3.5,-.3) node {\footnotesize$x$};
		\draw [->] (0,0) -- (0,6);
		\draw (-.3,6) node {\footnotesize$y$};
		\draw[thick] (2.5,5) arc [start angle=90, end angle=270, radius=2.5];
		\draw (.5,4.6) node {\footnotesize$\ms{NW}$};
		\draw (.5,.35) node {\footnotesize$\ms{SW}$};
		\draw (1.25,2.5) node {\footnotesize$\ms{D}$};
	\end{tikzpicture}
	\caption{Scaling limit of the rectangular lattice and the expected shape of the arctic curve in the free-fermion point. The regions $\ms{SW}$ and $\ms{NW}$ have ferroelectric ordering while the region $\ms{D}$ is disordered.}
	\label{res-lat}
\end{figure}
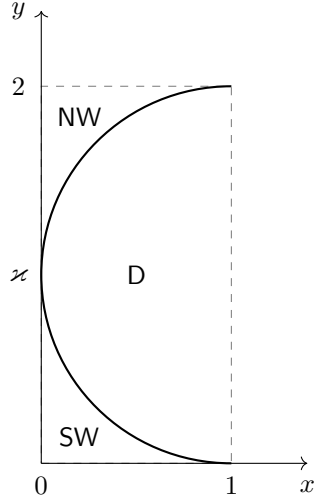

In the scaling limit, as we simultaneously take $N\rightarrow \infty$ 
and the spacing between vertices to zero, the lattice can be rescaled to a rectangle of dimensions $2\times 1$. We set the origin at the bottom-left corner of the lattice, as indicated in Figure \ref{res-lat}. 

In this setting, we define the scaling limit of $G_N^{(r)}$ as
\begin{align}
	G(y)=\lim_{r,N\rightarrow\infty}G_N^{(r)},&&y=\lim_{r,N\rightarrow\infty}\frac{2(N-r)}{N},&&y\in[0,2].
\end{align}
Recall the definition of $G_N^{(r)}$ is the probability of having a 
down-arrow  
between the $r$th and $(r+1)$th double rows. On the other hand, in the scaling limit of the disordered regime we expect the north-west ($\ms{NW}$) and south-west ($\ms{SW}$) regions of the lattice to be filled with $w_1$ and $w_2$, respectively. See Figures \ref{6v}, \ref{latt-rdwbc} and \ref{res-lat}. Consequently, the polarization state of all edges in the $N$th column from the bottom up to the contact point is $\ket{\downarrow}$, and $\ket{\uparrow}$ from this point forward. This means $G(y)$ has a stepwise behavior going abruptly from $1$ to $0$ once we cross the contact point at the left boundary. Denoting the contact point as $\vkappa$, we have
\begin{align}
	G(y)=\begin{cases}
	1,& 0\le y < \vkappa\\
	0,& \vkappa< y \le 2.
	\end{cases}
	\label{step}
\end{align}

We can capture the behavior \eqref{step} by analyzing the integral formula of $G_N^{(r)}$  by means of the saddle-point method. To do this, we first require the integrand of \eqref{int-g} to be cast in the form $e^{N\vtheta(z)}$, where
\begin{align}
	\vtheta(z)=\lim_{N\rightarrow\infty}\frac{1}{N}\left[\log h_N(z)-\log(z-1)-r\log z\right].
\end{align}
Thus the saddle-point equation reads
\begin{align}
	\vtheta'(z_0)=\frac{y-2}{2z_0}+v(z_0)=0,
	\label{sp-1}
\end{align}
where $z_0$ is the saddle-point and the function $v(z)$ is defined as
\begin{align}
	v(z)=\frac{\dd}{\dd z}\lim_{N\rightarrow\infty}\frac{\log h_N(z)}{N}.
	\label{aux}
\end{align}
 Since our only result regarding the asymptotic limit of $\log h_N(z)$ is \eqref{asymp-log-h}, where $z=\gamma(\omega)$ (which is given by \eqref{gam}), we will only consider real and positive values of $z$ in our analysis. 

Let $C_0$ be the counterclockwise oriented contour obtained by deforming $C$ in order to pass through the saddle-point. Depending on the range of values of $z_0$, the integral \eqref{int-g} is $0$ or $1$ by virtue of the residue theorem. More specifically, if $z_0<1$ there are no singularities in the region delimited by $C_0$ except for the pole of order $r$ at $z=0$. Thus, the values of the integral over $C$ and $C_0$ are equal in this case. Since the asymptotic value of the integral over $C_0$ is
\begin{align}
\oint_{C_0}e^{N\vtheta(z)}\dd z\approx\left[\frac{2\pi}{N|\vtheta''(z_0)|}\right]^{1/2}e^{N\vtheta(z_0)}\rightarrow 0, && N\rightarrow \infty,
\label{asymp-int}
\end{align}
we are left with $G(y)=0$ if $z_0<1$. However, if $z_0>1$ the deformed path necessarily includes the singularity at $z=1$. This means that the integral over $C_0$ is \eqref{int-g} plus a contribution from the residue at this simple pole. Therefore
\begin{align}
\oint_{C}\frac{h_N(z)}{(z-1)z^r}=-\oint_{C_1}\frac{h_N(z)}{(z-1)z^r}=-2\pi\im,
\end{align}
where $C_1$ is a counterclockwise oriented path including only the pole at $z=1$. Then, $G(y)=1$. Thus we are led to conclude that the stepwise behavior of $G(y)$ happens when  $z_0=1$. In terms of $z=\gamma(\omega)$, this implies $\omega_0=0$. Substituting \eqref{asymp-log-h} into \eqref{sp-1} and solving for $y=\vkappa$, we get the intermediate expression
\begin{align}
	\vkappa=2\left[1-(1-\sin(2\lambda)\tan\omega)\frac{\cos(\lambda+\omega)\sin(\lambda-\omega)}{\sin(2\lambda)\cos(2\omega)}\right]_{\omega=0},
\end{align}
which yields $\vkappa=1$ for any value of $\lambda$.

\section{The Tangent Method}\label{tangent}
In this section, we apply the Tangent Method \cite{TANGENT} for the six-vertex model with reflecting end boundary condition in order to obtain the arctic curve of the model at the special point $\Delta=0$, $\mu=0$, $a=b$. 

\begin{figure}[h!]
	\centering
	\begin{tikzpicture}[scale=.95,>=Stealth]
	
	\draw (-3,2.25) node {$w_1$};
	\draw (-3,0) -- (-3,2);
	\draw (-4,1) -- (-2,1);		
		
	\draw (-.5,2.25) node {$w_1$};			
	\draw (-.5,0) -- (-.5,2);
	\draw (-1.5,1) -- (.5,1);	
	\draw [line width=2pt] (-1.5,1) -- (-.6,1) -- (-.5,1.1) -- (-.5,2);
	\draw [line width=2pt] (-.5,0) -- (-.5,.9) -- (-.4,1) -- (.5,1);
	
	\draw (2,2.25) node {$w_2$};			
	\draw (2,0) -- (2,2);
	\draw (1,1) -- (3,1);
	\draw [line width=2pt] (2,0) -- (2,2);

	\draw (4.5, 2.25) node {$w_2$};	
	\draw (4.5,0) -- (4.5,2);
	\draw (3.5,1) -- (5.5,1);
	\draw [line width=2pt] (3.5,1) -- (5.5,1);

	\draw (7, 2.25) node {$w_3$};			
	\draw (7,0) -- (7,2);
	\draw (6,1) -- (8,1);	
	\draw [line width=2pt] (7,0) -- (7,.9) -- (7.1,1) -- (8,1);
	
	\draw (9.5, 2.25) node {$w_3$};		
	\draw (9.5,0) -- (9.5,2);
	\draw (8.5,1) -- (10.5,1);	
	\draw [line width=2pt] (8.5,1) -- (9.4,1) -- (9.5,1.1) -- (9.5,2);		
	
	\draw (1,-1) -- (1.5,-1);
	\draw (1,-2) -- (1.5,-2);	
	\draw (1.5,-2) arc [start angle=-90, end angle=90,radius=.5];
	\draw (2.5,-1.5) node {$\kappa_+$};
	\draw [line width=2pt] (1.5,-1) arc [start angle=90, end angle=0,radius=.5];
	\draw [line width=2pt] (1.5,-1) -- (1,-1);
	
	\draw (4,-1) -- (4.5,-1);
	\draw (4,-2) -- (4.5,-2);	
	\draw (4.5,-2) arc [start angle=-90, end angle=90,radius=.5];
	\draw (5.5,-1.5) node {$\kappa_-$};
	\draw [line width=2pt] (4.5,-2) arc [start angle=270, end angle=360,radius=.5];
	\draw [line width=2pt] (4,-2) -- (4.5,-2);
	\end{tikzpicture}
	\caption{Equivalent way of representing the Boltzmann weights of the six-vertex model and the reflecting end boundary.}
	\label{6v-lines}
\end{figure}
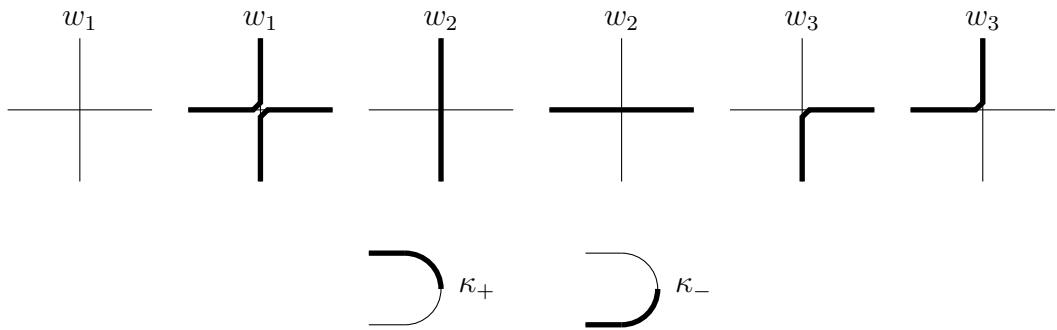

First of all, we recall there is an equivalent way to describe the six-vertex model other than we presented in Section \ref{sixvertex}: to each left or down arrow we place a thick line over the edge, and none otherwise. In this case the vertices are represented as in Figure \ref{6v-lines}.
In virtue of the ice-rule, the states are now characterized by non-crossing continuous lattice paths. These paths are directed in the sense that if we start to follow one of them rightward or upward there can never be a left or south step. It is worth noting that in the case of fixed boundary conditions all paths starting at the left side or bottom of the lattice must end at right or top boundaries. In fact, for the reflecting end boundary condition in a $2N\times N$ lattice (Figure \ref{latt-rdwbc}), all $N$ paths start at the bottom and end at the right boundary.

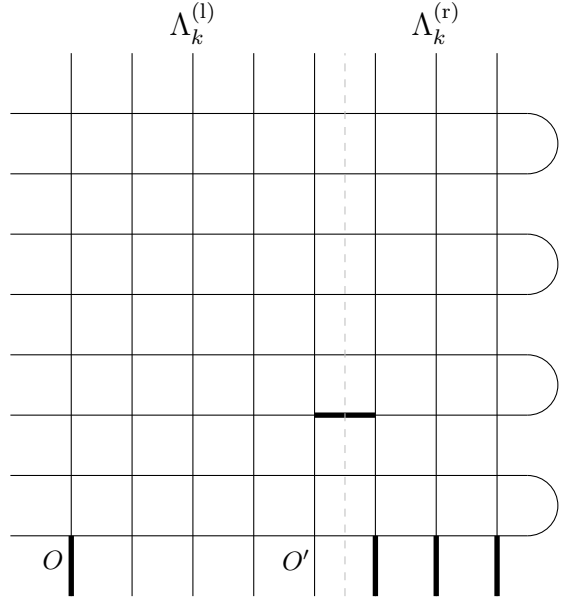
\begin{figure}[h!]
	\centering
	\begin{tikzpicture}[>=Stealth,scale=.8]
	
	\foreach \x in {-5,...,2}
	{  
		\draw (\x,0) -- (\x,9);
	}
	
	\foreach \x in {0,...,2}
	{
		\draw [line width=2pt] (\x,0) -- (\x,1);
	}
	
	\foreach \x in {1,...,8}
	{
		\draw (-6,\x) -- (2.5,\x);
	}
	
	\draw [line width=2pt] (-5,0) -- (-5,1);
	\draw [line width=2pt] (-1,3) -- (0,3);
	
	\draw [style=lightgray,dashed] (-.5,0) -- (-.5,9);
	
	\foreach \x in {1,3,5,7}
	{		
		\draw (2.5,\x) arc [start angle=-90,end angle=90, radius=.5cm];
	}
	
	\draw (-5.3,.6) node {\footnotesize$O$};
	\draw (-1.3,.6) node {\footnotesize$O'$};
	
	\draw (-3,9.5) node {$\Lambda_k^{(\mathrm{l})}$};
	\draw (1,9.5) node {$\Lambda_k^{(\mathrm{r})}$};
	\end{tikzpicture}
	\caption{The extended lattice with $N=L=4$ and $k=3$.}
	\label{extended}
\end{figure}

Consider a rectangular lattice with $N$ double rows and $N+L$ columns, as depicted in Figure \ref{extended}, with $L\in \mathbb{N}$. This lattice can be seen as the juxtaposition of two domains, namely
namely $\Lambda_k^{(\mathrm{r})}$ and $\Lambda_k^{(\mathrm{l})}$, $k=1,\ldots,2N$, with dimensions $2N\times(N-1)$ and $2N\times (L+1)$ respectively. We assign thick lines for the $(N-1)$ last bottom vertical edges, as well as for the leftmost one, and thin lines for the remaining boundary edges. 

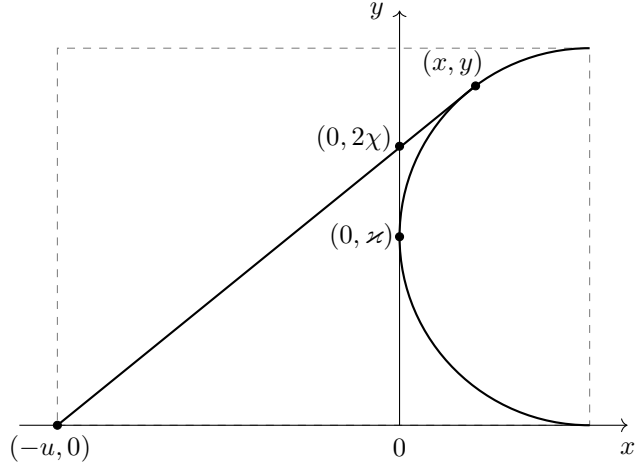
\begin{figure}[h!]
	\centering
	\begin{tikzpicture}
		
		\draw[color=black!50,dashed] (-4.5,0) rectangle (2.5,5);
		\draw (0,-.3) node {\footnotesize$0$};
		\draw (-.5,2.5) node {\footnotesize$(0,\vkappa)$};
		\draw [->] (-5,0) -- (3,0);
		\draw (3,-.3) node {\footnotesize$x$};
		\draw [->] (0,0) -- (0,5.5);
		\draw (-.3,5.5) node {\footnotesize$y$};
		\draw [thick] (-4.5,0) -- (1,4.5);
		\draw [fill=black] (1,4.5) circle [radius=1.5pt];
		\draw [fill=black] (0,3.7) circle [radius=1.5pt];
		\draw [fill=black] (-4.5,0) circle [radius=1.5pt];
		\draw [fill=black] (0,2.5) circle [radius=1.5pt];
		\draw (-4.6,-.3) node {\footnotesize$(-u,0)$};
		\draw (-.6,3.8) node {\footnotesize$(0,2\chi)$};
		\draw  (.7,4.8) node {\footnotesize$(x,y)$};
		\draw [thick] (2.5,5) arc [start angle=90, end angle=270, radius=2.5];
	\end{tikzpicture}	
	\caption{Scaling limit of the extended lattice.}
	\label{extended-sl}
\end{figure}

Let us analyze these domains individually, starting with $\Lambda_k^{(\mathrm{l})}$. There is only one thick edge at the bottom of this domain. Since all the edges on its left and top boundaries are fixed, this thick edge originates a directed path that starts one north step before the origin $O=(0,0)$ and reaches the interface between the two domains at $(L,k-1)$. As for the domain $\Lambda_k^{(\mathrm{r})}$, there are $N$ paths ending at the right boundary, $N-1$ of which start at the south-side plus one additional path entering its left boundary. In this setting, we apply the Tangency Assumption \cite{TANGENT} and state that in the scaling limit: 
\begin{enumerate}
	\item The $N-1$ paths leaving the bottom of the domain $\Lambda_k^{(\mathrm{r})}$ give rise to the same disordered region as the $2N\times N$ lattice with reflecting end boundary.
	\item The directed path in $\Lambda_k^{(\mathrm{l})}$ becomes a straight line that crosses the boundary between the domains at $(0,k/N>\vkappa)$ (with respect to $O'$) and it is tangent to the north-west portion of the arctic curve at $(x,y)$. Since paths do not cross, from the tangency point up to the top contact point the additional path is expected to bend to the shape of the arctic curve, and then go straight until it reaches the right boundary. 
\end{enumerate}
See Figure \ref{extended-sl} for a sketch of these assumptions. We stress the additional path shall not make an angle when crossing from the left to the right domain since the majority of Boltzmann weights in $\Lambda_k^{(\mathrm{l})}$ are the same as the ones in the frozen corner outside the arctic curve in $\Lambda_k^{(\mathrm{r})}$ above the contact point $\vkappa$.

Then, the north-west part of arctic curve is the envelope of the family of straight lines $U_{u}(x,y;z)=0$ obtained by varying the parameter of extension of the original lattice, $L=u N$. The arctic curve parametric coordinates $x(z)$, $y(z)$ are solutions of the system of equations
\begin{align}
	U_{u}(x,y;z)=0,&&\frac{\dd}{\dd z}U_{u}(x,y;z)=0.
\end{align}

Let $k=2\chi N$. With respect to the origin $O'$, the equation for a straight line that crosses the $x$ axis at $(-u,0)$ and the $y$ axis at $(0,2\chi)$ reads 
\begin{align}
	U_{u}(x,y;z)=y-\frac{2\chi}{u}x-2\chi=0.
	\label{eq-sl}
\end{align}
Thus, the problem of finding the arctic curve is reduced to the determination of parameters $\chi$ and $u$. This is done by computing asymptotic behavior of the partition function $Z_{N,L}$ of the extended lattice, which is given by
\begin{align}
	Z_{N,L}=\sum_{k=1}^{2N}Z_k^{(\mathrm{l})}Z_k^{(\mathrm{r})},
	\label{part-nl}
\end{align}
where $Z_k^{(\mathrm{l,r})}$ are the partition functions of $\Lambda_k^{(\mathrm{l,r})}$ domains.

Let us first consider $Z_k^{(\mathrm{l})}$. The configurations on this domain are the possible ways one can reach $(L,k-1)$ from $O=(0,0)$ through a directed lattice path with north and east steps only. Of course, the enumeration of paths must take into account the different contributions of vertices $w_2$ and $w_3$. In the following, we proceed along the same lines as \cite{TANGENT}.

In order to count the paths, we weight them according to their number $\ell$ of east-north corners (vertices preceded by an east step and followed by a north step).  If $\ms{w}$ is the contribution of each of these corners, the weighted enumeration of paths starting from $(0,0)$ and reaching $(x,y)$ is the sum
\begin{align}
	P_{\ms{w}}(x,y)=\sum_{\ell\ge 0}\mc{N}(x,y;\ell)\ms{w}^{\ell},&&\mc{N}(x,y;\ell)=\binom{x}{\ell}\binom{y}{\ell},
\end{align}
where $\mc{N}(x,y;\ell)$ is the number of paths from $(0,0)$ to $(x,y)$ with $\ell\le\min\{x,y\}$ east-north corners. Is it important to note that the for a given path $\Gamma:(0,0)\rightarrow(x,y)$, the path $\widetilde{\Gamma}$ obtained from the former by adding a north step just before the origin and an east step right after the final point has the same number of east-north corners, i.e. $\ell(\Gamma)=\ell(\widetilde{\Gamma})$. 

Instead of assigning the weight $\ms{w}$ to east-north corners we can consider two different weights: $\ms{w}_1$ for when the steps surrounding a vertex are of the same kind, and $\ms{w}_2$ otherwise. Let $s(\widetilde{\Gamma})$ and $t(\widetilde{\Gamma})$ be the number of $\ms{w}_1$ and $\ms{w}_2$ vertices, respectively, on a given path $\wt{\Gamma}$. One can relate $\ell(\wt{\Gamma})$ to these two quantities as
\begin{align}
	s(\widetilde{\Gamma})+t(\widetilde{\Gamma})=x+y+1,&&t(\widetilde{\Gamma})=2\ell(\widetilde{\Gamma})+1.
\end{align}
From Figure \ref{6v-lines} we see that it is plausible to assign the weight of vertices of type $s$ and $t$ to $w_2/w_1$ and $w_3/w_1$, respectively. Generally speaking, the problem of weighted enumeration of paths in the six-vertex model with reflecting end boundary can be quite complicated because not only there is a change in sign in the spectral parameter every other line, but also because the reading of vertices replaces $a_-\rightarrow b_+$, $b_-\rightarrow a_+$ when going from the bottom to the top line of a given double row. Nevertheless, in the case $\Delta=\mu=0$ we can choose $\lambda=\pi/4$ so $a=b=1/\sqrt{2}$ and $c=1$. In this setting, the alluded distinction between steps on successive horizontal lines no longer applies and we can simply use the result derived for the case of domain wall boundary, namely \cite{TANGENT}

\begin{align}
	P_{a}(x,y)=\sum_{\ell\ge 0}\binom{x}{\ell}\binom{y}{\ell}a^{-(2l+1)}.
	\label{enum-dwbc}	
\end{align}
From the above result we obtain the partition function by multiplying \eqref{enum-dwbc} by the normalization factor $w_1^{2N(L+1)}$, which gives
\begin{align}
	Z_{k}^{(\mathrm{l})}=\sum_{\ell\ge 0}\binom{L}{\ell}\binom{k-1}{\ell}a^{2N(L+1)-2\ell-1}.
	\label{part-left}
\end{align}

As for the domain $\Lambda_k^{(\mathrm{r})}$, we see this is the $2N\times N$ original lattice with reflecting end boundary after excluding the $N$th column in which there is a $c$-vertex at the $k$th row (from the bottom). Therefore this partition function is related to the correlation $H_N^{(r)}$, $r=1,\ldots,N$. In fact, it is clear from Figure \ref{a-d} that in the general case
\begin{align}
	Z_{k=2n}^{(\mathrm{r})}&=\frac{A_N^{(N-n+1)}(\lambda,\mu)}{cb_-(\lambda,\mu)(a_-(\lambda,\mu)b_+(\lambda,\mu))^{N-n}(a_+(\lambda,\mu)b_-(\lambda,\mu))^{n-1}},\label{part-a}\\
	Z_{k=2n-1}^{(\mathrm{r})}&=\frac{D_N^{(N-n+1)}(\lambda,\mu)}{cb_+(\lambda,\mu)(a_-(\lambda,\mu)b_+(\lambda,\mu))^{N-n}(a_+(\lambda,\mu)b_-(\lambda,\mu))^{n-1}}
	\label{part-d}	
\end{align}
which simplify to
\begin{align}
	Z_{k=2n}^{(\mathrm{r})}=\frac{A_N^{(N-n+1)}}{a^{2N-1}},&&Z_{k=2n-1}^{(\mathrm{r})}=\frac{D_N^{(N-n+1)}}{a^{2N-1}},
	\label{part-right}
\end{align}
when $\mu=0$, $a=b$, $c=1$.

Substituting \eqref{part-left} and \eqref{part-right} in \eqref{part-nl} we obtain
\begin{align}
	Z_{N,L}=a^{2NL}\sum_{n=1}^{N}\sum_{\ell\ge 0}a^{-2\ell}\binom{L}{\ell}\left[\binom{2n-1}{\ell}A_N^{(N-n+1)}+\binom{2n-2}{\ell}D_N^{(N-n+1)}\right],
\end{align}
where we changed the sum index $k\rightarrow n$ when separating the even and odd terms. In terms of $H_N^{(N-n+1)}$,
\begin{align}
	Z_{N,L}=a^{2NL} Z_N \sum_{n=1}^{N}\sum_{\ell\ge 0} a^{-2\ell}\binom{L}{\ell}\binom{2n-1}{\ell}H_N^{(N-n+1)}\left[1-\frac{\ell}{2n-1}\frac{D_N^{(N-n+1)}}{Z_N H_N^{(N-n+1)}}\right].
	\label{z-nl}
\end{align}

We are now ready to consider the scaling limit. We transform the sums in \eqref{z-nl} into integrals in the variables $\chi=n/N$ and $\zeta=\ell/N$ and apply the saddle-point method in order to find the asymptotic behavior of $Z_{N,L}$. Thus
\begin{align}
	\widetilde{Z}_{N,L}=\frac{Z_{N,L}}{a^{2NL}Z_N}\propto \int {\cal S}(\xi,\zeta;u)\dd\chi\dd\zeta,
\end{align} 
where  
\begin{align}
	&{\cal S}(\chi,\zeta;u)=2\chi\log 2\chi-2\zeta\log(a\zeta)-(2\chi-\zeta)\log(2\chi-\zeta)+u\log u-\nonumber \\
	&-(u-\zeta)\log(u-\zeta)+\lim_{N\rightarrow\infty}\frac{1}{N}\left(\log H_N^{(N-n+1)}+\log\left[1-\frac{u}{2\chi}\frac{D_N^{(N-n+1)}}{Z_N H_N^{(N-n+1)}}\right]\right).
	\label{integrand}
\end{align}
By definition, $Z_N H_N^{(r)}=A_N^{(r)}+D_N^{(r)}\gtrsim D_N^{(r)}$ since both terms are positive, which makes the argument of the logarithm in the last term of \eqref{integrand} to fall within the interval $(0,1)$. Therefore, its contribution can be neglected in the limit $N\rightarrow\infty$. Imposing $\del {\cal S}/\del\chi|_{\chi_0,\zeta_0}=0$, $\del {\cal S}/\del\zeta|_{\chi_0,\zeta_0}=0$, we are left with
\begin{align}
	0&=2\log\left[\frac{2\chi_0}{2\chi_0-\zeta_0}\right]+\frac{\dd}{\dd \chi}\lim_{N\rightarrow\infty}\frac{1}{N}\log H_N^{(N-n+1)}\Bigg{|}_{\chi=\chi_0},\label{chi} \\
	0&=\log\left[\frac{2(2\chi_0-\zeta_0)(u-\zeta_0)}{\zeta_0^2}\right],\label{zeta}
\end{align}
where we used $a=1/\sqrt{2}$. In order to find the asymptotic behavior of $H_N^{(N-n+1)}$ we turn to the generating function $h_N(z)$. In fact, in the scaling limit we can rewrite the definition \eqref{gen} as
\begin{align}
	h_N(z)\propto \int e^{Np(\chi)}\dd\chi,&&p(\chi)=\lim_{N\rightarrow\infty}\frac{1}{N}\log H_N^{(N-n+1)}+(1-\chi)\log z.
	\label{gen-sl}
\end{align}
The major contribution to the integrand comes from the maximum point of $p(\chi)$. Imposing $p'(\chi)=0$ we arrive at
\begin{align}
	\frac{\dd}{\dd\chi}\lim_{N\rightarrow\infty}\frac{1}{N}\log H_N^{(N-n+1)}\Bigg{|}_{\chi_0}=\log z.
	\label{log-h-z}
\end{align}
Note that $p''(\chi)=(\dd/\dd\chi)^2\log H_N^{(N-n+1)}$, so the sign of the second derivative of $p(\chi)$ is the same as $\log H_N^{(r)}$. As a function of $\chi$, the correlation $H_N^{(r)}$ assumes values within the range $(0,1]$, which makes its logarithm to be a concave function of $\chi$ with maximum attained at the contact point. We are led to conclude that $p''(\chi_0)<0$ and therefore $\chi_0$ is a maximum point, as desired. 

With the result $\eqref{log-h-z}$ we are now able to solve the system \eqref{chi}-\eqref{zeta}, which gives
\begin{align}
	\zeta_0^{(\pm)}=\pm\frac{2u\sqrt{z}}{1\pm\sqrt{z}},&&\chi_0^{(\pm)}=\pm\frac{u\sqrt{z}}{1-z},
	\label{solutions}
\end{align}
with equal signs to be taken simultaneously. Note that we must choose the pair of solutions such that $\zeta_0$ and $\chi_0$ are both positive, which depends on whether $z>1$ or not. If we take $z=\gamma(\omega)$ (given by \eqref{gam} with $\lambda=\pi/4$), then $z\in(0,+\infty)$ as $\omega\in(-\pi/4,\pi/4)$ with $z=1$ at $\omega=0$. Therefore the adequate solutions are
\begin{align}
	(\chi_0,\zeta_0)=\begin{cases}
	\left(\chi_0^{(+)},\zeta_0^{(+)}\right),&\omega\in(-\pi/4,0)\\
	\left(\chi_0^{(-)},\zeta_0^{(-)}\right),&\omega\in(0,\pi/4)
	\end{cases}.
	\label{poss}
\end{align}
On the other hand, we can obtain the saddle-point $\chi_0$ by taking the derivative of asymptotic value of the integral \eqref{gen-sl} with respect to $z$. Indeed, in this situation we have $h_N(z)\propto e^{Np(\chi_0)}$ and therefore
\begin{align}
	v(z)=\frac{\dd p(\chi_0)}{\dd z}&&\Longrightarrow&&\chi_0=1-z\ v(z),
	\label{saddle-chi}
\end{align}
where $v(z)$ is given by \eqref{aux}. Substituting the slope $\chi_0/u$ \eqref{poss} and $\chi_0$ \eqref{saddle-chi} into \eqref{eq-sl}, we obtain two possibilities for the function $U_u(x,y;z)$, each corresponding to a different envelope of straight lines depending on the interval of $z$. In order to obtain the north-west portion of the arctic curve, we must choose $z<1$, which yields
\begin{align}
	U_u(x,y;z)=y-\frac{2\sqrt{z}}{1-z}\ x-2(1-v(z)).
\end{align}
Finally, imposing $U_u=(\dd/\dd z)U_u=0$ and solving for $x(z),\ y(z)$, we arrive at the parametric coordinates of the north-west portion of the arctic curve
\begin{align}
	x(z)&=\dfrac{2\sqrt{z}(1-z)^2[v(z)+zv'(z)]}{1+z},\label{xz}\\
	y(z)&=\dfrac{2(1+z)+2z(1-3z)v(z)+4z^2(1-z)v'(z)}{1+z}\label{yz},
&&z\in(0,1).	
\end{align}
In terms of $\omega$, expressions \eqref{xz} and \eqref{yz} can be brought into
a more intuitive form
\begin{align}
	x(\omega)=1-\cos(2\omega),&&y(\omega)=1-\sin(2\omega),&&\omega\in\Big(-\frac{\pi}{4},0\Big).
	\label{arc-nw}
\end{align}
We can obtain the south-west portion of the curve from this result. In fact, note that the $\ms{NW}$ and $\ms{SW}$ regions are filled with $w_1$ and $w_2$ weights, respectively, which differ by the orientation of the arrows in their vertical edges. Thus, the south-west portion of the curve is obtained by the transformations $\mu\rightarrow-\mu$, $x\rightarrow x$, $y\rightarrow 2-y$ in \eqref{arc-nw} and therefore the whole curve is a semicircle centered at $(1,1)$ with unit radius (see Figure \ref{res-lat}). It is worth noting this result is in agreement with Monte Carlo simulations \cite{LYBERG} and coincides with the west portion of the
arctic curve of the six-vertex model with domain wall boundary conditions in the $2N\times 2N$ square lattice \cite{ARTIC}.

\section{Conclusion}\label{conclusion}

In this paper we analytically computed the arctic curve for the six-vertex model with reflecting end boundary condition in the special point $\Delta=0,\ \mu=0,\ a=b$ by means of the Tangent Method \cite{TANGENT}. In order to do this, we used the boundary correlations introduced in \cite{PASSOS}, where the essential ingredient is the generating function $h_N(z)$. This function is directly related to these correlations and consequently appears in the algebraic equations leading to the contact point with the left boundary as well as to the parametric equations for a portion of the curve. The remaining part of the curve is obtained by symmetry relation. The arctic curve in this case is shown to be a semicircle, which is in good agreement with numerical results \cite{LYBERG}.

Nevertheless, we have not determined the arctic curve in the general case $a\ne b$. This is due to the fact that the weighted enumeration of directed paths in the rectangular lattice becomes much more challenging since the weights of vertices along the paths may change depending on the horizontal row that they are placed.
In addition,  we have not dealt with the cases of  $\Delta\neq 0$ and $\mu\neq 0$. This is due to limitations we encounter in the evaluation of the asymptotic behavior of the generating function $h_N(z)$. The main problem is due to the nature of the one-point boundary correlation function which in the case of the reflecting end boundary is given in terms of two pieces \eqref{corr-h}. The difficulties were circumvented only in the case of $\Delta=0$ and $\mu=0$, which made the expression more symmetric and therefore  manageable. In conclusion, we still need to generalize the current methods to determine the arctic curve for the general case $\Delta\neq 0$, $\mu\neq 0$ and $a\neq b$. We hope to address such problems  in the future.

\section*{Acknowledgments}

The authors thank the S\~ao Paulo Research Foundation (FAPESP) for financial support through the grants 2017/22363-9 and 2017/16535-1.

\end{document}